\documentstyle[11pt]{article}

\begin{document}
\large

\def\lsim{\mathrel{\rlap{\lower3pt\hbox{\hskip0pt$\sim$}}
    \raise1pt\hbox{$<$}}}         
\def\gsim{\mathrel{\rlap{\lower4pt\hbox{\hskip1pt$\sim$}}
    \raise1pt\hbox{$>$}}}         
\def\dblint{\mathop{\rlap{\hbox{$\displaystyle\!\int\!\!\!\!\!\int$}
}
    \hbox{$\bigcirc$}}}
\def\ut#1{$\underline{\smash{\vphantom{y}\hbox{#1}}}$}

\newcommand{\beq}{\begin{equation}}
\newcommand{\eeq}{\end{equation}}
\newcommand{\dem}{\Delta M_{\mbox{B-M}}}
\newcommand{\dega}{\Delta \Gamma_{\mbox{B-M}}}

\newcommand{\ind}[1]{_{\begin{small}\mbox{#1}\end{small}}}
\newcommand{\tind}[1]{^{\begin{small}\mbox{#1}\end{small}}}

\newcommand{\WA}{{\em WA}}
\newcommand{\SM}{Standard Model }
\newcommand{\QCD}{{\em QCD }}
\newcommand{\KM}{{\em KM }}
\newcommand{\hscale}{\mu\ind{hadr}}
\newcommand{\sG}{i\sigma G}

\newcommand{\MS}{\overline{\mbox{MS}}}
\newcommand{\pole}{\mbox{pole}}
\newcommand{\aver}[1]{\langle #1\rangle}

\newcommand{\appa}{\mbox{\ae}}
\newcommand{\CP}{{\em CP } }
\newcommand{\fy}{\varphi}
\newcommand{\hi}{\chi}
\newcommand{\al}{\alpha}
\newcommand{\as}{\alpha_s}
\newcommand{\gf}{\gamma_5}
\newcommand{\gm}{\gamma_\mu}
\newcommand{\gn}{\gamma_\nu}
\newcommand{\be}{\beta}
\newcommand{\ga}{\gamma}
\newcommand{\de}{\delta}
\renewcommand{\Im}{\mbox{Im}\,}
\renewcommand{\Re}{\mbox{Re}\,}
\newcommand{\GeV}{\,\mbox{GeV}}
\newcommand{\MeV}{\,\mbox{MeV}}
\newcommand{\matel}[3]{\langle #1|#2|#3\rangle}
\newcommand{\state}[1]{|#1\rangle}
\newcommand{\ra}{\rightarrow}
\newcommand{\ve}[1]{\vec{\bf #1}}

\newcommand{\rhs}{{\em rhs}}
\newcommand{\pp}{\langle \ve{p}^2 \rangle}

\newcommand{\BR}{\,\mbox{BR}}
\newcommand{\La}{\overline{\Lambda}}
\newcommand{\Lam}{\Lambda\ind{QCD}}

\newcommand{\ACM}{AC$^2$M$^2$ }

\newcommand{\np}{non-perturbative }
\newcommand{\re}[1]{Ref.~\cite{#1}}

\begin{flushright}
\large{
CERN-TH.7159/94\\
TPI-MINN-94/2-T\\
UMN-TH-1235-94}\\
UND-HEP-94-BIG02\\
$[$hep-ph/9402225$]$\\
February 1994\\
\end{flushright}
\vspace{.4cm}
\begin{center} \LARGE
{\bf Heavy Quark Distribution Function in QCD and
the AC$^2$M$^2$ Model}
\end{center}
\vspace*{.4cm}
\begin{center} {\Large
I. Bigi $^{a,b}$, M. Shifman $^c$, N. Uraltsev $^{a,d}$,
A. Vainshtein $^{c,e}$} \\
\vspace{.4cm}
{\normalsize $^a${\it TH Division, CERN, CH-1211 Geneva 23,
Switzerland}\footnote{During the academic year 1993/94}\\
$^b${\it Dept.of Physics,
University of Notre Dame du
Lac, Notre Dame, IN 46556, U.S.A.}\footnote{Permanent address}\\
$^c$ {\it  Theoretical Physics Institute, Univ. of Minnesota,
Minneapolis, MN 55455}\\
$^d$ {\it St.Petersburg Nuclear Physics Institute,
Gatchina, St.Petersburg 188350, Russia}$^2$\\
$^e$ {\it Budker Institute of Nuclear Physics, Novosibirsk
630090, Russia}\\
\vspace{.4cm}
e-mail addresses:\\
{\it BIGI@CERNVM, SHIFMAN@VX.CIS.UMN.EDU,
VAINSHTE@VX.CIS.UMN.EDU}
}
\end{center}
\thispagestyle{empty} \vspace{1.4cm}

\centerline{\Large\bf Abstract}
\vspace{.4cm}
We show that the phenomenological \ACM ansatz is consistent with QCD
through order $1/m_b$ in the description of
$B\ra l\bar \nu_l+X_u$ and $B\ra \gamma +X_s$ transitions, including
their energy spectra and differential distributions.
This suggests a concrete realization for the QCD distribution
function, which we call the ``Roman'' function. On the other hand
the \ACM model description of the end-point domain in
$B\ra l\bar \nu_l + X_c$ is incompatible with QCD: a different distribution
function enters the description of $b\ra c$ decays as compared to the
transitions to the massless quarks. Both observations -- the validity of
the {\ACM}-like description for
heavy-to-light transitions and the emergence of the new distribution
function in the $b\ra c$ case -- are in
contradiction to a recent claim in the literature.
The intrinsic limitation of the \ACM
model could reveal itself in different values of the effective $b$ quark
mass from fits of the $\Lambda _b$ and $B$ decays.

\newpage
\large
\addtocounter{footnote}{-2}

1. In recent publications \cite{1,2} we have shown how
the lepton and the photon energy spectra
in semileptonic and radiative decays, respectively,
of beauty hadrons
can be predicted
from QCD proper without recourse to phenomenological
{\em ad hoc} assumptions; this is achieved by employing a heavy
quark expansion in inverse powers of the beauty quark mass
to deal with \np  corrections.
It turned out that the shape of the spectrum in the
end-point region requires a special treatment (see also
Refs. \cite{3,4}). The physical phenomenon behind that is quite
transparent: the heavy
quark $Q$ moves inside the decaying hadron $H_Q$ and this motion smears
the lepton
spectrum, most significantly in the end-point region.
It has been noted \cite{1,2} that a rigorous QCD
description -- though having a similar origin --
does not generically coincide with simple non-relativistic
models of the `Fermi' motion where the potential energy of
the light degrees of
freedom is always small as compared to their space-like momentum; the most
popular model of this type
is due to \ACM \cite{5}. Original generalities of the QCD approach are
given in Refs.~\cite{9,old,8}.

The \ACM model describes the beauty meson $B$
in very simple terms. It treats it as consisting of the
heavy $b$ quark plus a spectator with fixed mass $m_{sp}$; the latter
is of the order of a typical hadronic scale and represents a
fit parameter. The spectator quark has a momentum distribution
$\Phi (|\vec{p}|)$ where $\vec p$ is its three-dimensional
momentum.
The $b$ quark then cannot possess a
fixed mass. Indeed, the energy-momentum conservation implies for such
a simple boundstate that the $b$ quark energy $E_b$
is given by $M_B-\sqrt{|\vec p|^2+m_{sp}^2}$. The mass
of the $b$ quark equals $E_b$ through order $p$; thus one
obtains a ``floating'' $b$ mass
\beq
m_b^f\simeq  M_B-\sqrt{|\vec p|^2+m_{sp}^2}
\label{1}
\eeq
which depends on $|\vec p|$.
The lepton spectrum is first obtained from $b$ quark decay in a moving
frame where its momentum is $-\vec p$; the quark mass is assumed to be
$m_b^f$.  Then the distribution is averaged with
the weight function $\Phi (|\vec{p}|)$ (further details can be found in
the original work \cite{5}). The Galilean transformation above
introduces a Doppler smearing of the spectrum that is
of order $1/m_b$; the transverse Doppler shift arises
only at order $1/m_b^2$ and is ignored here.

The  AC$^2$M$^2$ model is  extensively
used in the analysis of the lepton energy spectrum in
semi-leptonic decays. Surprising though it is,  given all
the na\"{i}vet\`{e} of the  model, it  reproduces very well numerically
the {\em shape} of the
semi-leptonic spectra  in its regular part, as it is derived  from QCD
(see \cite{1}).
In this paper we analyze in more detail what features
of QCD are reflected in  this model for the end-point spectrum.  We will show
that -- contrary to a
first impression -- this
model is built sufficiently well to reproduce correctly the
pattern of the QCD description of the heavy quark motion
provided one considers the transition of the heavy to a
{\em massless} quark.
Moreover, the specific choice adopted in \ACM for the weight
function $\Phi (|\vec{p}|)$ \cite{5} corresponds to a particular form of the
genuine distribution function $F(x)$ introduced in Ref. \cite{2} which
schematically can be viewed as

\beq
\Phi(|\vec{p}|) =\frac{4}{p_F^3\sqrt{\pi}} {\rm e}^{-
\frac{\vec{p}^2}{p_F^2}}\;\;\;
\longleftrightarrow  \;\;\;\Phi(p_{+})=\theta (p_{+} )
\frac{1}{p_F\sqrt{\pi}}{\rm e}^{-\frac{(p_{+}-\frac{m_{sp}^2}{p_{+}})
^2}{4p_F^2}}
\eeq
where $p_+$ is the light cone combination $p_0+p_z$ entering
the QCD description
(the exact relation to $F(x)$ will be stated below). This represents
an acceptable realization of the QCD distribution function meeting all
necessary requirements (we will call it the Roman function
\footnote{The possible shape of the function shown in Figs.~1,2
demonstrates that it is definitely not of a {\sf sans serif} type.});
it can easily
accommodate  improvements should they become necessary.
At the same time the \ACM model {\em per se} does not reproduce
correctly the dependence on the final state quark mass $m_q$ if the mass
is large enough. As follows from the QCD analysis \cite{2} the boundary
lies at $m_q\sim (\La m_b)^{1/2}$ and $c$ quark is sufficiently heavy
in this sense. Moreover, in the  $b\ra c$ transitions
the \ACM model in its original form would yield spectra beyond
the physical kinematic  boundary. A good fit of the experimental lepton
energy spectrum in $b\ra c$ is achieved \cite{1} at a price of introducing an
upper cut off in the heavy quark momentum depending on the final
quark mass.

Recently a comparison of the QCD result and of the \ACM
model has been given in Refs. \cite{6,7},
where it was pointed out
that the lepton energy spectrum obtained from the
\ACM ansatz can be considered to be consistent with the
QCD prediction in the sense that there are no
corrections to order $1/m_b$. In this note we
give a more detailed analysis. We also point out that the consideration
of Refs.~\cite{3,4} in the part concerning the \ACM model is incorrect.

\vspace*{.4cm}

2. Some of the dynamical features in the spectra of charged
leptons get obscured by the integration over the neutrino
energy. They are seen in a cleaner way in the spectrum of photons in
the radiative $B\ra \gamma +X_s$ transitions.
This spectrum in the free quark
approximation  is just
a monochromatic line:
\beq
\frac{d\Gamma^{(0)}}{dE} =\frac{\lambda^2}{4\pi} m_b^3\delta (E-
\frac{m_b}{2})
\label{3}
\eeq
where $E$ is the photon energy, $m_b$ is the (current) $b$ quark
mass, the strange quark is considered as massless and $\lambda$ is
the coupling constant determining the
$bs\gamma$ vertex as it emerges from the electroweak theory:
\beq
{\cal L}\vert_{b\rightarrow s\gamma} =
\frac{i}{2}\lambda F_{\mu\nu}\bar s (1+
\gamma_5)\sigma_{\mu\nu} b .
\label{2}
\eeq
In both the QCD approach \cite{1,2,4} and in the \ACM
model \cite{5} the monochromatic line of eq. (\ref{3}) is
transformed into a
peak of a finite width due to the heavy quark motion.
The width of the peak is of the order of a typical hadronic scale.
Note, that in the \ACM model the variation of the $b$ quark mass
(see below) also contributes to the width.

Gluon radiation will also lead to a smearing out of the photon line.
Since our primary object
here is to discuss non-perturbative effects like
the Fermi motion, we will ignore
perturbative gluon emission throughout most of this paper.

Non-perturbative corrections are incorporated in the \ACM model
through the three fit parameters $m_{sp}$, $m_q$
and $p_F$. Let us recall now some general expressions derived
previously \cite{2} from QCD for $m_q=0$ case.  Neglecting the gluon
radiative corrections one can write the spectrum for the inclusive
radiative decay in the form
\beq
\frac{d\Gamma}{dE} =\frac{\lambda^2}{4\pi} m_b^3
\frac{2}{\bar \Lambda}F(x)
\label{A}
\eeq
where $\La = M_B- m_b$,
\beq
x=\frac{1}{\bar \Lambda} (2E-m_b) \;\;,
\label{B}
\eeq
and $F(x)$ is the QCD distribution function. The moments of $F(x)$ ,
\beq
a_n=\int dx x^n F(x) \;\;,\;\;\;  n =0,1,...
\label{C}
\eeq
are related to the expectation values of local
operators taken between $B$ mesons:
\beq
\frac{1}{2M_B}\langle B|\;{\cal S}\;\,\bar b
\pi_{\mu_1}...\pi_{\mu_n}b-\;
\mbox{traces }|B\rangle =
 a_n{\bar \Lambda}^n (v_{\mu_1}...v_{\mu_n} -\; \mbox{traces}).
\label{D}
\eeq
Here ${\cal S}$ is the symmetrization symbol and $v_\mu$ is the
four-velocity of the $B$ meson. Moreover,
\beq
\pi_\mu = iD_\mu -m_bv_\mu
\label{E}
\eeq
where $D_\mu$ is the covariant derivative.
Note that $a_0=1$ and $a_1=0$
(these relations hold up to $1/m_b^2$ corrections which we
disregard throughout). An important consequence of quantum
chromodynamics is that the total width
\beq
\Gamma =\frac{\lambda^2}{4\pi} m_b^3+...
\label{F}
\eeq
receives no corrections at the level $1/m_b$; they arise first
at order $1/m_b^2$ \cite{old,8}.

It should be noted that the true distribution function $F(x)$
is always one-dimen\-si\-o\-nal.
For in the QCD description
different components of the momentum represented by
$\pi_\mu$ do not commute and the distribution function
can be introduced only for a single
combination of
the components of the momentum. In the case at hand, when the
final quark is massless, i.e. $m_q=0$,
it is the `light cone' distribution function that
enters (i.e. we actually deal with the distribution in $\pi_0 +\pi_z$
where the z-axis is chosen along the direction of the photon
momentum; for more details see Ref. \cite{2}).

At first sight the two approaches -- that of Refs.
\cite{1,2,4} on the one hand, and  the
AC$^2$M$^2$ model \cite{5} on the other -- have completely
different properties.
Two main distinctions  are obvious.

The   AC$^2$M$^2$ model introduces  the  distribution over
the {\em three-dimensional }
nonrelativistic momentum, $\Phi (p) $ where $p =|\vec p |$ is the
absolute value of the $b$ momentum. This distribution results in  the
Doppler
smearing of the photon spectrum
with the spread  {\em linear} in $1/m_b$.
Moreover,  the AC$^2$M$^2$ prescription
replaces $m_b$ by $m_b^f$, the floating quark mass, see
eq. (\ref{1}):
\beq
\frac{d\Gamma_{ACM}}{dE} =\frac{\lambda^2}{8\pi}
\int p^2 dp\, dz \:\Phi (p)\,(m_b^f)^3\delta (E-
\frac{m_b^f}{2}-\frac{pz}{2})\;\;,
\label{4}
 \eeq
where $z=\cos\theta$ and $\theta$ is the angle between $\vec p$ and
the direction of the photon momentum. (Eq. (\ref{4}) is obtained by
substituting $m_b$ by $m_b^f\,$, making the
Lorentz boost and then
convoluting the spectrum resulting from the parton  approximation with
$\Phi (p)$.)

At the same time,
the QCD expansion operates in terms of the fixed, current mass of the
the $b$
quark, $m_b$. The effect of the Fermi motion enters via the
light cone distribution function defined in eqs. (\ref{C}), (\ref{D})
(we remind the reader that   the
final $s$ quark is treated as massless like the $q$ quark before).
\vspace*{0.4cm}

3. We will show now that in spite of these differences the two approaches
can actually lead to the same photon spectrum.
The \ACM ansatz yields for the photon spectrum
after averaging over the direction of $\vec p$
\beq
\frac{d\Gamma_{ACM}}{dE} =\frac{\lambda^2}{8\pi}
\theta (\epsilon )
\int_{p_0^2(\epsilon )}^\infty \, dp^2 \Phi (p^2)
\left( M_B-\sqrt{m_{sp}^2+p^2}\, \right)^3,
\label{5}
\eeq
with
\beq
\epsilon = M_B - 2E\;\;, \;\;\; \; \;
p_0^2(\epsilon ) = \frac{1}{4}\left( \frac{m_{sp}^2}{\epsilon}-
\epsilon\right)^2 .
\label{7}
\eeq
The spectrum
is automatically cut off at the true kinematic boundary, i.e.
at $E=M_B/2$ by the step function for $\epsilon <0 $,
irrespective of the
particular choice of $\Phi (p)$. This is not surprising, of course,
because the adoption of the floating $b$ quark mass
was dictated just by this requirement.

The shape of the spectrum is obtained by
direct integration and depends on the specific
form of $\Phi (p)$. The distribution $\Phi (p)$ suggested in Ref. \cite{5} and
routinely
used in experimental analyses is
\beq
\Phi (p)=\frac{4}{p_F^3\sqrt{\pi}} {\rm e}^{-\frac{p^2}{p_F^2}}
\label{8}
\eeq
with the following normalization
$$
\int_0^\infty dp\, p^2 \Phi (p) = 1.
$$
Notice that the upper limit of integration (which must be $M_B$) is
extended to $\infty$. This introduces only an exponentially small
error; below we will
always use this extension of the upper limit of integration.

Using the ansatz of eq.~(\ref{8})
and keeping terms through order $p_F/M_B$ one gets:
\beq
\frac{d\Gamma_{ACM}}{dE}=
\left(\frac{\lambda^2 M_B^3}{4\pi}\right)\theta (\epsilon )
\frac{2}{p_F\sqrt{\pi}}{\rm e}^{-\frac{p_0^2(\epsilon )}{p_F^2}}
\left[ 1-\frac{3p_F}{M_B}{\rm e}^{(\rho +\frac{p_0^2
(\epsilon )}{p_F^2})}\Gamma (\frac{3}{2},\rho
+\frac{ p_0^2 (\epsilon )}{p_F^2} )
+{\cal O}(\frac{p_F^2}{M_B^2})\right]
\label{9}
\eeq
where the energy $E$ enters via $\epsilon$, $p_0^2(\epsilon )$ defined in
eq.~(\ref{7}) and
\beq
\rho =\frac{m_{sp}^2}{p_F^2}\;\; .
\label{10}
\eeq
The factor in the square brackets can be put to unity in
discussing the
{\em shape} of the spectrum in this approximation. The term
proportional to $p_F/M_B$ includes the incomplete
gamma function $\Gamma(\alpha ,x)$ and will be taken into account below to
obtain
the total width to $1/m_b$ accuracy.
The expression in eq.~(\ref{9}) exhibits a pronounced
peak whose shape depends, of course,
on the value of $\rho$ (see Fig.$\;$1). For $\rho <1$ it is rather
asymmetric, a second gratifying feature of the \ACM model,
since it is in qualitative accord with the findings in QCD. As
we will see later on, this model can provide an approximate
description of the real world only when $\rho$ is rather small.

Straightforward integration of the spectrum eq. (\ref{5}) leads to
\beq
\Gamma_{ACM} =\frac{\lambda^2M_B^3}{4\pi}
\left[ 1-\frac{3p_F\rho}{M_B\sqrt{\pi}} {\rm e}^{\rho /2}
K_1(\frac{\rho}{2}) +{\cal O}(\frac{p_F^2}{M_B^2})\right]
\label{11}
\eeq
where $K_1$ is the McDonald function.
Once $\Gamma _{ACM}$ is expressed in terms of the quantity
\beq
m_b^{ACM} = M_B -\frac{p_F\rho}{\sqrt{\pi}}{\rm e}^{\rho /2}
K_1(\frac{\rho}{2})
\label{12}
\eeq
(which, as we will see in a moment, is
nothing but the value of the floating mass $m_b^f$
of eq. (\ref{1}) averaged over the
distribution $\Phi(p)\,$),
the correction to first order in $1/m_b$ is eliminated from the
total width,
\beq
\Gamma_{ACM} =\frac{\lambda^2(m_b^{ACM})^3}{4\pi}
\left( 1+{\cal O}(1/m_b^2)\right),
\label{13}
\eeq
in full agreement with the general statement of the absence of the
$1/m_b$ correction in the total width \cite{old,1,8}.
Thus, one must identify
\beq
M_B-m_b=\La =  \frac{p_F\rho}{\sqrt{\pi}}{\rm e}^{\rho /2}
K_1(\frac{\rho}{2}) \ra\left\{
\begin{array}{ll}
2p_F/\sqrt{\pi} & \mbox{at $\rho\ra 0$} \\
m_{sp}  & \mbox{at  $\rho\ra \infty$}
\end{array}
\right.
{}.
\label{14}
\eeq

That the first order correction to the total width can be
absorbed into model parameters is not
surprising. What is less obvious
is that this definition is compatible with QCD predictions for other
quantities as well.
We will show now that this is indeed the case. (The same
conclusion concerning inclusive semi-leptonic decays
has been reached previously  in Refs.~\cite{6,7}.)

{}From the QCD analysis of Refs. \cite{2,21} we
know the average photon energy that points to the true mass $m_b$ :
\beq
\langle\epsilon\rangle = \langle M_B -2E \rangle
=\La (1 - \langle  x \rangle ) =\La
\label{15}
\eeq
where $x$ is defined by  eq. (\ref{B}) and $\langle  x \rangle
=a_1 =0$.
To find $\langle  \epsilon \rangle$ in the \ACM model we
use eq.~(\ref{5}) or eq.~(\ref{9}).
Now the difference between the quark
mass and $M_B$
can be ignored and one obtains
\beq
\langle\epsilon\rangle _{ACM}=
\frac{p_F\rho}{\sqrt{\pi}}{\rm e}^{\rho /2}
K_1(\frac{\rho}{2}) \left (1+{\cal
O}(1/m_b)\right) .
\label{17}
\eeq
which agrees with the QCD expression eq.~(\ref{15}) provided that the
identification of eq.~(\ref{14}) is made.  Moreover, this holds true
for any choice of $\Phi (p)$.

For the second moment of the photon energy one derives from
eq. (\ref{5})
\beq
\langle\epsilon^ 2\rangle _{ACM} = p_F^2 (2+\rho ).
\label{18}
\eeq
On the other hand, according to Ref. \cite{2}
\beq
\langle\epsilon ^2\rangle ={\La}^2 (1+ \langle x^2\rangle ) ,
\label{19}
\eeq
where $\langle x^2\rangle$ is related to the average kinetic energy
of the $b$ quark inside the $B$ meson, see eqs. (56) and (94) in
\cite{2},
\beq
\langle x^2\rangle = \frac{1}{3\La^2}(2M_{B})^{-1}
\langle B|\bar b{\vec\pi}^2 b|B\rangle
\equiv\frac{\mu_\pi^2}{3\La^2} .
\label{20}
\eeq
Comparing eqs.~(\ref{18}) and (\ref{19}) we conclude that
\beq
p_F^2 (2+\rho ) =\La^2 (1+ \langle x^2\rangle ) =
\frac{p_F^2\rho^2}{\pi}{\rm e}^{\rho}
K_1^2(\frac{\rho}{2})
(1+ \langle x^2\rangle ) .
\label{21}
\eeq
Assuming that $\langle x^2\rangle $ is known from a QCD analysis --
and to a certain extent  this is indeed the case -- we can use
eq.~(\ref{21}) to fix the value of $\rho$ which then may serve as an
input
in eq. (\ref{14})  allowing one to determine $p_F$.
Moreover, rewriting eq. (\ref{21}) in the form
\beq
1+ \langle x^2\rangle =\pi (2+\rho )\rho^{-2}{\rm e}^{-\rho}
K_1^{-2}(\frac{\rho}{2})
\label{G}
\eeq
it is easy to see that the right-hand side is a monotonously
decreasing function of $\rho$ having its maximum ($\approx 1.6$)
at $\rho = 0$. Thus, in the \ACM model an upper bound emerges,
\beq
\langle x^2\rangle\; < \frac {\pi}{2} -1 \simeq 0.57
\label{U}
\eeq
On the other hand,
the value of
$\mu_\pi^2$ has been estimated \cite{10} from QCD sum rules
\cite{11} to be  $\mu_{\pi}^2 \sim 0.6$ GeV$^2$ and $\La \simeq
0.4\div 0.6 \GeV$.
Taking these estimates at face value one would conclude that
$\langle x^2\rangle \sim 0.5\div 1$. This would mean that
the \ACM ansatz can be made consistent with the real QCD
description only for small values of $\rho$, i.e. when the
spectator is relativistic. Further
numerical estimates will be presented below.

The higher moments are obtained in a straightforward way,
\beq
\langle\epsilon^n\rangle_{ACM} = \frac{p_F^n}{\sqrt{\pi}}
\rho^{\frac{n+1}{2}}{\rm e}^{\frac{\rho}{2}} K_{\frac{n+1}{2}}
(\frac{\rho}{2}) \;\,.
\label{H}
\eeq
For even values of $n$ the McDonald function in the right-hand side
reduces to an elementary one,
\beq
\langle\epsilon^{2k}\rangle =p_F^{2k}\sum_{l=0}^k
\frac{(2k-l)!}{l!(k-l)!}\rho^l\;\; .
\label{I}
\eeq
For any value of $n$ the limits of small and large $\rho$ are
\beq
\langle\epsilon^n\rangle\ra
\left\{
\begin{array}{ll}
2^n p_F^n \Gamma (\frac{n+1}{2})/\sqrt{\pi} & \mbox{at
$\rho\ra 0$} \\
p_F^n\rho^{n/2}(1+\frac{n(n+2)}{4\rho})=m_{sp}^n
(1+\frac{n(n+2)}{4}\frac{p_F^2}{m_{sp}^2})
 & \mbox{at  $\rho\ra \infty$}
\end{array}
\right.
{}.
\label{K}
\eeq
The moments $\langle\epsilon ^n\rangle$ given in  eq. (\ref{H}) are
related to those of an `equivalent' \ACM distribution function
{\em via} the following expression
\beq
\langle\epsilon ^n\rangle ={\bar \Lambda}^n
\langle (1-x)^n\rangle \;\;.
\label{M}
\eeq
In particular
combining eq.~(\ref{H}) for $n=3$ and eq.~(\ref{G}) we can express the third
moment,
$\aver{-x^3}$, via another monotonously
decreasing function of $\rho$ implying that
\beq
0 < - \langle x^3 \rangle < 2- \frac{\pi}{2}\simeq 0.43\;\;;
\label{N}
\eeq
for small $\rho$ the upper bounds of eqs.~(\ref{U}), (\ref{N})
actually become approximate equalities. One should note
that the QCD estimate \cite{2} (see also \cite{mannel}),
$$\aver{-x^3}\approx 0.03\GeV^3/\La^3 \sim 0.3\div 0.5 \; ,$$
is then quite compatible with the \ACM value.
\vspace*{0.4cm}

4. The equivalent `light cone' distribution function $F(x)$ -- the Roman
function --
can be read
off from eq.~(\ref{9})
taking into account the
definition of the scaling variable $\epsilon =\La (1-x ) $,
\beq
F_{Rom}(x)= \frac{1}{\sqrt{\pi}} \frac{\La}{p_F}
\exp \left\{ -\frac{1}{4}\left[
\frac{p_F}{\La}\frac{\rho }{1-x} -
\frac{\La}{p_F} (1-x)\right]^2\right\}\;\;,
\label{22}
\eeq
\nopagebreak
$$
\frac{\La}{p_F}=\frac{\rho}{\sqrt{\pi}}{\rm e}^{\rho /2}
K_1(\frac{\rho}{2})\;\;.
$$
This expression represents how the \ACM ansatz
can be reproduced by a QCD light cone function.
The distribution given by the Roman function of
eq.~(\ref{22}) satisfies all requirements
necessary for the
generic distribution function governing inclusive  decays into
massless quarks. As was mentioned, it
does not
lead to any spectrum beyond the kinematical boundary $M_B/2$ and
it exponentially
decreases towards large negative $x$. Of course the latter property for
$F(x)$ must be
considered as a goal in the mature QCD description incorporating
radiative corrections rather than an obvious property;
on the other hand it is
most natural to require it from the model function introduced to describe
non-perturbative effects.

If $m_{sp}\gg
p_F$ the Roman function $F_{Rom}$ exhibits a
narrow peak around $x\simeq 0$ and
represents the situation when even the spectator is nonrelativistic.
The opposite limiting case is more
relevant, $p_F\gsim m_{sp}$. In this case $F_{Rom}$ is rather broad --
its width is $\sim\La$ -- and very asymmetric, see Fig.~1.

\vspace*{0.4cm}

5. Next we extend our analysis to
inclusive semi-leptonic $B$ decays.
The question which we want to address here
is supplementary to the one
treated in Refs. \cite{6,7}.  Namely, we are interested in studying
the scaling features of the differential distribution
$$
d\Gamma (B\rightarrow l \bar{\nu}_l X_u)/dE_l dq^2 dq_0
$$
where $E_l$ is the charged lepton energy and $q_\mu=(p_l +p_\nu
)_\mu$ is the total four-momentum of the lepton pair.  The analysis
of Ref. \cite {2} implies that
\beq
m_b^5 \;\frac{d\Gamma (B\rightarrow l \bar{\nu}_l X_u)}{dE_l dq^2
dq_0}\;=\;
\Gamma_0 \:(b\rightarrow l \bar{\nu}_l\, u) \,\frac{2}{\bar\Lambda}
F(x)\,
\frac{12(q_0 - E_l)(2m_bE_l - q^2)}{(m_b - q_0)}
\label{23}
\eeq
where
\beq
\Gamma_0 (b\rightarrow l \bar{\nu}_l\, u)= |V_{ub}|^2
\frac{G_{F}^2 m_b^5}{192\pi^3} .
\label{24}
\eeq
In other words, we predict that  the differential distribution -- which
generically is a function of two independent variables, $q_0$ and
$q^2$ -- actually depends only on a single scaling combination,
\beq
x =-\frac{m_b^2+q^2-2m_bq_0}{2\La (m_b-
q_0)}
\label{25}
\eeq
apart from a simple kinematical factor.  Like in deep inelastic scattering,
this scaling holds only when both
perturbative and higher twist effects
are neglected. Both effects introduce corrections to
the scaling regime which are not discussed here.

We would like to check whether this scaling feature -- the
dependence on a single scaling variable -- persists in the \ACM
model.  To this end we just repeat the steps carried out above
for the radiative transition, see eq.~(\ref{5}). After averaging over
the direction of $\vec p$, the primordial momentum
of the $b$ quark, we
get for
$d^3\Gamma /dE_ldq^2dq_0$ an integral representation quite similar
to eq. (\ref{5}). What is different is the lower limit $p_0^2$ of
integration over $p^2$ which now takes the form
\beq
p_0^2 =\frac{1}{4}\left( \frac{m_{sp}^2}{\tilde\epsilon}-
\tilde\epsilon\right)^2 ,
\label{26}
\eeq
\beq
\tilde\epsilon =
M_B\frac{M_B^2+q^2-2q_0M_B}{M_B^2-q^2} .
\label{27}
\eeq
This expression is sufficient to see that the dependence on $q^2$ and
$q_0$ enters only {\em via} the variable $\tilde\epsilon$.

Moreover, $\tilde\epsilon$ is directly expressed in terms of the
QCD variable $x$ of eq. (\ref{25}),
\beq
\tilde\epsilon = \La (1-x) \left[ 1+{\cal O}(\frac{\La}{m_b}(1-
x))\right] .
\label{28}
\eeq
The \ACM model, thus, reproduces the correct QCD
scaling taking place  in the leading approximation.

\vspace*{0.4cm}

6. There is, however, an important
feature in the true QCD description
that is beyond the scope of the \ACM ansatz:
this model, as it is formulated in \cite{5}, is supposed
to be universally applicable to the case of massless
and of massive quarks in the final state, i.e. to
$b\ra u$  and to $b\ra c$ transitions, respectively.
On the other hand, we have shown \cite{2} that the genuine
distribution function depends in an essential way on the
ratio of the quark masses in the initial
and the final state,
$$
F=F(x,\ga)\;\;,\;\;\ga=m_q^2/m_b^2.
$$
The function is strikingly different in the two extreme
cases when $\ga\ll 1$ and $1-\ga\ll 1$; the latter represents the
so-called
small velocity (SV) limit \cite{SV}.  Let us demonstrate that in this
limit the \ACM model would lead to predictions which
differ essentially from the true QCD result.

In the SV limit the velocity of the final quark $q$ is small, $|\vec v
|<<1$; this can happen both in radiative transitions and in
semi-leptonic decays; in the latter case it requires
$$0<(m_Q-m_q)^2-q^2\ll m_q m_Q\;\;\;.$$
and in the former
$$
\Delta m = m_Q -m_q << m_Q .
$$
When one retains only terms
${\cal O}(v^0)\,$, the physical spectrum in the
$B$ meson decay is
exactly the same as it was at the free quark level -- the $\de$-function
peak resides at the same
place, and
the inclusive probability is completely saturated by one heavy
meson in
the final state \cite{SV}.
Notice that there is no `$M_B-m_b$ window' -- no shift
is present between the
maximal allowed energies of the photon (or lepton)  at the quark and
the hadron
levels.

Modifications of this perfect quark-hadron duality
start at the level of ${\cal O}(v^2)$ \cite{bjorken2}.  If ${\cal O}(v^2)$
effects are considered
 the height of the elastic peak is changed, and a comb of
inelastic peaks appears, the height of the latter being proportional to
$v^2$. This comb will lie at $E<E_0$ and will be stretched over an
energy interval of order $\La$. ( $\La$ is not a very relevant
parameter in this limit;
we can continue to use it, however, just as a typical hadronic scale.
One could certainly choose another definition of the typical hadronic
scale.) The integral over the inelastic peaks must compensate the
distortion of the elastic one --  the so-called Bjorken sum rule
\cite{bjorken2}. In the recent paper \cite{2} we demonstrated how this
apparently different description emerges from the same QCD analysis that
leads to the {\ACM}-like prediction for massless final-state quarks.

Let us compare this picture to the one from the \ACM
model. Doppler smearing yields again one smooth peak whose width
is now of order $\La v$. The \ACM prescription thus produces nothing
that resembles the two-component picture outlined above.
The failure of this description has a
clear origin -- it incorporates Fermi motion in the initial state but
disregards it in the final state where it is now crucial.

For the same reason in the actual
applications of the model in $b\ra c$ transitions one has to introduce an
{\em ad hoc} upper cut off for the heavy quark momentum, as was
mentioned in the Introduction. This cut off is determined by the physical
kinematic boundary and is, thus, non-universal, i.e. depends on the
final quark mass.

The energy spectrum in the SV limit of QCD
is given by a `temporal' distribution
function having in general a discrete support representing the higher excited
states and supplemented by continuum contributions with at least two extra
pions in the final state.
One then has an Isgur-Wise -like \cite{IW} description of the spectrum where
instead of relying on the QCD expressions for inclusive widths one is
suggested to sum explicitly over the particular final state hadrons. To
leading order in $|\vec{v}|^2$ one then has only $D^*\,(D)$ and
in
the next order adds higher excitations \footnote{This is similar to a
modification of the original I-W ansatz, incorporating $D^{**}$ states,
as now used
in experimental analyses of semileptonic spectra.}. Adopting this
description one then can fit the spectrum by a different
phenomenological function not directly related to the Fermi motion; it
is important however that in the framework of
the QCD expansion there are certain relations between the moments of the
Fermi motion distribution function and the parameters of the temporal one
(cf. \re{2}). These relations must be observed in comparing decay spectra with
massless and with heavy quarks in the final state.

The mass of the actual $c$ quark is such that one
might think at first sight
it could be safely treated as massless since $m_c^2/m_b^2
\sim 0.1$.  This naive expectation seems to be erroneous, though.
For the true parameter that enters
in semi-leptonic decays is rather given by the
ratio
$$ \tilde{\ga}=\frac{4m_b^2m_c^2}{(m_b^2+m_c^2-q^2)^2}$$
and its value is typically not very small even in the upper end
of the charged lepton energy once it is integrated over the invariant mass
squared of the lepton pair $q^2$.
The conjecture that the $c$ quark actually lies rather close to the SV limit
is supported by the observation that the hadronic final state in $B$ meson
semi-leptonic decays consists, to large degree, of only a
$D$ or a $D^*$.  (In the exact
limit, semileptonic $B$ decays would produce only
a $D$ or a $D^*$ \cite{SV}.)
If so, it is clear that the attempts to fit the parameters of the \ACM
model directly from the data on the inclusive $b\ra c$ transitions,
a quite popular procedure, are not very meaningful from a truly
theoretical perspective.
One can further get convinced that this is indeed the case
by estimating the value of $m_c$ as it emerges from these fits --
it turns out to be too high from any standpoint.

It has been claimed in \re{3}
that the same description in terms of shape
functions holds for both $b\ra u$ and $b\ra
c$ semileptonic transitions, and that the same hadronic matrix elements
define it for an arbitrary ratio $m_c^2/m_b^2$. It is clear from
the analysis of \re{2} that such claims are erroneous.

\vspace*{.4cm}

7. The intrinsic inadequacy of the \ACM ansatz for describing
semileptonic
$b\ra c$ transitions in the SV limit gets obscured once one
integrates over the neutrino momentum to obtain the charged
lepton energy spectrum. Fitting $B\ra l+X_c$ by the \ACM model one
has commonly set $m_{sp}$ to 150 MeV in a more or less
ad-hoc fashion; the charm mass together with $p_F$ are then used
as free fit parameters. In an attempt to give this model a somewhat
closer connection to QCD one should actually adopt a different
strategy: namely to require that the difference between the
\ACM average of the $b$ quark mass and the charm quark mass satisfy
the same relationship with
heavy flavor hadron masses that
the heavy quark expansion yields in QCD for $m_b-m_c$, i.e.
\beq
\aver {m_b}_{ACM}-m_c\simeq \frac{3M_B^*+M_B}{4}-\frac{3M_D^*+M_D}{4}
+ \aver{\vec{\pi}^2}
\frac{m_b-m_c}{2m_bm_c}
\eeq
Instead one allows $m_{sp}$ together with $p_F$ to
float in the fit. It is quite conceivable that such an approach
would yield values for the fit parameters in better
agreement with QCD expectations. We do not have at hand the
results of such an
analysis; still,
to have an example of the possible distribution function one can use
the existing
fit parameters from CLEO data \cite{fit}, namely $p_F\simeq
282\MeV$ and
$m\ind{sp}\simeq 150\MeV$.
{}From eqs.~(\ref{14}-\ref{15}, \ref{H}, \ref{M}) we get
$$\La\simeq  360 \MeV\;\;,\;\; \aver{\vec{\pi}^2}\simeq
0.16\GeV^2$$
\beq
a_2\simeq 0.424\;,\;\;\;a_3\simeq -0.28\;,\;\;\;a_4\simeq 0.71\;,\;\;\;
a_5\simeq -1.2
\label{40}
\eeq
and the distribution function itself in terms of the photon energy is  shown in
Fig.~2. Keep in mind that this curve should be taken only as an illustration
since the perturbative
gluon corrections, very important for the numerical analysis, are not
included here.

One should note that the value of the kinetic energy operator is slightly
below the lower
bound derived in \re{2}. This points again to the intrinsic inconsistency of
the
literal \ACM approach with semileptonic decays into charm. The third moment
of the
distribution functions is somewhat less than the estimate of \re{2} as well.
All these facts seem
to be correlated with the too high mass of the $c$ quark obtained in the
\ACM fit of such decays.
In reality one can obtain better information on the
distribution function $F(x)$ at small $\ga$ studying the double
differential
distributions and in particular in the region of small $q^2$
\cite{koyrakh,2}.
It is worth reminding however that according to the analysis of \re{1} the
calculable $1/m_Q^2$ corrections are really important for the fit and
they
must be taken into account.
\vspace*{.4cm}

8. So far we have disregarded gluon  radiative corrections that actually
modify the end point
shape in an essential way; their interplay with the effects of the Fermi motion
has been discussed
in ref.~\cite{2}. In radiative decays of the type $b\ra s + \ga $ the peak in
the photon energy is still manifest although its height is
significantly
lowered by gluon bremsstrahlung. Both perturbative and \np effects smearing
the
original narrow line can be described with sufficient practical accuracy
by convoluting the radiatively corrected parton spectrum
with the distribution function describing Fermi motion \cite{2}; the radiative
corrections must be considered in higher orders of perturbative expansion as
well, at least in the double logarithmic approximation.
A more precise treatment
including subleading perturbative corrections requires further theoretical
consideration. We note that it is just this theoretical prescription that has
been employed in the existing phenomenological analysis (for radiative
decays,
see \re{Ali}.). In semileptonic decays integrating over the neutrino energy
smooths out the spectrum and the impact of radiative corrections on the shape
is
less pronounced.

The conjecture in Refs.~\cite{3,4} that radiative corrections produce
relatively small, namely ${\cal O}(\alpha_s (m_b))$, modifications to the
moments of the
end-point distribution functions is certainly incorrect.

\vspace*{0.4cm}

9. In the conclusion of this paper we would like to add some
comments on existing
comparisons of the predictions from the \ACM model with
 QCD . The most obvious remark follows
from the fact that there are no corrections of order
$1/m_b$ to total {\em widths}, as stated first
in Refs.~\cite{old,8};
thus, for example, differences in the lifetimes of $B$ and $\Lambda_b$ arise
first
on the $1/m_b^2$ level; the corrections due to Fermi motion are
quadratic in
$1/m_b$ as well. This does not hold automatically in \ACM;
yet by defining an effective $b$ quark mass $m_b^{ACM}$ as the
average over the floating mass of eq.~(1) one eliminates $1/m_b$ corrections
to both the total width and to the regular part of the lepton spectrum.
This observation
has been noted in a few
recent papers \cite{6,7}. The above analysis shows that the similarity
extends also to the subtle problem of the endpoint description for decays into
light quarks --
contrary to
superficial claims in Refs.~\cite{3,4}. As a matter of fact the corresponding
considerations  of Refs.~\cite{3,4} do not refer to the \ACM model at all since
the crucial effect of the floating mass $m_b^f$ has been omitted. As a result
the kinematical boundaries were violated and the numerical values of the
moments obtained there are irrelevant.
In contrast, the analysis of $b\ra s+\ga$ in Ref.~\cite{Ali} has been
done properly: e.g. it preserved the kinematic constraints.

Of course essential differences
arise at order $1/m_Q^2$; however one should keep in mind that
the main
point of the \ACM model was to take care of only leading $1/m_Q$
effects
that are crucial even in beauty particles.

In our opinion, this is actually not the main difference,
probably not even from a practical point of view.
For it is important to understand that the value of the
$b$ quark mass $m_b^{ACM}$ defined in such a way can actually appear
to be {\em
different} for different beauty hadrons $H_b$ and even for the
different
channels ($b\ra c$ vs. $b\ra u$ or $b\ra s+\ga$) in the same hadron.
The
latter concern follows from the fact that
the spectrum is shaped by different distribution functions depending
on the final state quark mass; only the second moments of
these distribution functions coincide. This difference
can actually lead to the emergence of
different values for $m_b^{ACM}$ beyond the nonrelativistic
approximation
$p_F\ll m_{sp}\,$.

Once semileptonic or radiative decays of beauty baryons are studied,
the second concern arises:
it is more than conceivable that the literal
use of
the \ACM fit for the decays of baryons and mesons would yield
values of
$m_b^{ACM}$ that differ already in $1/m_Q$ terms; following the standard
\ACM
rules for calculating the width one then would arrive at
different
lifetimes already in the leading non-perturbative approximation. For the
average mass of the heavy quark plays only a marginal role in the model, and
thus essentially depends on the assumed shape of the distribution. The whole
distribution function $F(x)$ can be determined from a precise measurement of
the end point spectrum, in particular in $b\ra s+\ga$ decays. This of
course would yield the true $ b$ quark mass. On the other hand lepton spectra
in semi-leptonic decays are less sensitive to the exact shape and thus
effectively rely more on the model assumptions.
For in general there is no reason to expect that the particular \ACM
ansatz for the distribution function holds with a good accuracy for real
boundstates in the presence of potential-like binding forces. For example,
a possible and very natural generalization of the ansatz eq.~(\ref{1})
would be to
add some arbitrary ``static'' binding energy $V$ of the order of $\Lam\,$:
\beq
M_B\simeq  m_b^f+\sqrt{|\vec p|^2+m_{sp}^2}+V
\label{last}
\eeq
where $V$ may differ from hadron to hadron. This modification
cannot be absorbed into a redefinition of $m_{sp}$ and $p_F$.
No model can actually provide us with an unambiguous prescription for
determining the impact of $V$; QCD on the other hand ensures that
the effect of $V$ on the total width is cancelled by an analogous
interaction in the final state.
Therefore it seems to be
advantageous to impose additional constraint on the fitted distribution
functions ensuring them to yield the same QCD value of the `average' mass
$m_b$ independently of a particular ansatz used for their shape.

We think that the possibility to obtain different mass values  for the
same heavy quark
is actually the main {\em physical} indication of the
inconsistency of the standard \ACM ansatz with QCD, together with  a
difference in the
underlying Fermi motion description for different final state quark
masses -- which in the case of the SV limit is in a sense a rather
obvious observation.

\vspace*{0.5cm}

{\bf ACKNOWLEDGEMENTS:} \hspace{.4em}
We thank G.~Altarelli for his thoughtful comments on the manuscript.
N.U. acknowledges interesting discussions
with T.~Mannel and is grateful to D.~Cassel for his interest. M.S. and A.V. are
grateful to M.~Voloshin for useful comments.
We are also indebted to R. Mountain for producing the figures.
This work
was supported in part by the National Science Foundation under the
grant number PHY 92-13313 and by DOE under the grant
number DOE-AC02-83ER40105.

\vspace*{0.5cm}

{\bf Figure captions}\\

\vspace{0.3cm}

\noindent Fig. 1. The QCD distribution function $F(x)$ from the \ACM model
(Roman function) for $\rho=1$
(solid line), $\rho=0.2$ (dotted line) and $\rho=5$ (dashed line);$\;\;
\rho=m_{sp}^2/p_F^2\;$.
\\

\noindent Fig. 2. The distribution over the light cone momentum $\pi_0
+\pi_z$
corresponding to the \ACM model with $p_F=282\MeV$ and
$m_{sp}=150\MeV\,$.
\\


\begin{thebibliography}{99}

\bibitem{1}
I. Bigi, M. Shifman, N. Uraltsev and A. Vainshtein,  {\it Phys. Rev.
Lett}, {\bf 71}  (1993) 496.

\bibitem{2}
I. Bigi, M. Shifman, N. Uraltsev and A. Vainshtein,
{\bf preprint}  CERN-TH.7129/93, 1993.

\bibitem{3}
M. Neubert ,  {\bf preprint}  CERN-TH.7087/93, 1993.

\bibitem{4}
M. Neubert ,  {\bf preprint}  CERN-TH.7113/93,  1993.

\bibitem{5}
G. Altarelli et al., Nucl. Phys. B208 (1982) 365.

\bibitem{9}
J. Chay, H. Georgi and B. Grinstein, {\it Phys. Lett.} {\bf B247} (1990)
399.

\bibitem{old}
I. Bigi, N. Uraltsev and A. Vainshtein, {\it Phys. Lett.} {\bf B293}
(1992)
430; (E) B297 (1993) 477;\\
B. Blok and M. Shifman, {\it Nucl. Phys.} {\bf B399} (1993) 441; 459;

\bibitem{8}
I. Bigi, B. Blok, M. Shifman, N. Uraltsev and A. Vainshtein, {\it Proc. of
the
1992 DPF meeting of APS}, Fermilab, November 1992  [Preprint
UND-HEP-92-BIG07].

\bibitem{6}
C. Csaki and L. Randall, {\bf preprint} CTP-2262, 1993.

\bibitem{7}
G. Baillie, {\bf preprint} UCLA/93/TEP/47.

\bibitem{21}
A. Falk, M.Luke, M.Savage, {\bf preprint} UCSD/PTH 93-23.

\bibitem{10}
P. Ball and V. Braun, {\bf Preprint} MPI-Ph/93-51, 1993.

\bibitem{11}
For a review see
M. Shifman, Ed., {\em Vacuum Structure and QCD Sum Rules},
North-Holland, 1992.

\bibitem{mannel}
T.Mannel, private communication; {\it in preparation}.

\bibitem{SV}
M. Voloshin and M. Shifman, {\it Yad. Fiz.} {\bf 47} (1988) 801
[{\it Sov. J. Nucl. Phys.} {\bf 47} (1988) 511].

\bibitem{bjorken2}
J. Bjorken, Invited Talk at {\it Les Rencontres de la Valle d'Aosta,
La Thuille, 1990} Preprint SLAC-PUB-5278, 1990;\\
see also N. Isgur and M. Wise, {\it Phys. Rev.}  {\bf D43} (1991) 819.

\bibitem{IW}
B.Grinstein, N.Isgur, M.B.Wise, {\it Phys.Rev.Lett.} {\bf 56} (1986) 258.

\bibitem{fit}
J.Bartelt et al. (CLEO collab.) ``Inclusive Measurement of B Meson
Semileptonic Branching Fractions'', Cornell Univ. Preprint CLEO CONF 93-19.
1993.

\bibitem{koyrakh}
B. Blok, L. Koyrakh, M. Shifman and A. Vainshtein, Preprint TPI-
MINN-93/33-T [Phys. Rev. D, to be published];\\
A. Manohar and M. Wise, Preprint UCSD/PTH 93-14;\\
T. Mannel, Preprint IKDA 93/26.

\bibitem{Ali}
A.Ali, C.Greub, {\it Zeit. Phys.} {\bf C49} (1991) 431; {\it Phys. Lett.} {\bf
B259} (1991) 182; for recent update, {\it Zeit. Phys.} {\bf C60} (1993)
433.



\end{thebibliography}
\end{document}